\documentclass[aps,pre,reprint,amsmath,amssymb]{revtex4-1}
\usepackage[utf8]{inputenc}
\usepackage[pdftex]{graphicx}
\usepackage{graphicx,subfigure}
\usepackage{dcolumn}
\usepackage{epstopdf}
\usepackage{color}
\usepackage{upgreek}
\usepackage[hidelinks]{hyperref}

\usepackage{bm}

\makeatletter
\def\btt#1{\texttt{\@backslashchar#1}}%
\DeclareRobustCommand\bblash{\btt{\@backslashchar}}%
\makeatother

\begin{document}
\title{Smectic-like rheology and pseudo-layer compression elastic constant of a twist-bend nematic liquid crystal}

\date{\today}
\author{M. Praveen Kumar$^{1}$, P. Kula$^{2}$,  and Surajit Dhara$^{1}$}
\email{sdsp@uohyd.ernet.in} 
\affiliation{$^1$School of Physics, University of Hyderabad, Hyderabad-500046, India\\
$^{2}$Institute of Chemistry, Faculty of Advanced Technologies and Chemistry, Military University of Technology, Warsaw, Poland. }

\begin{abstract}
 In twist-bend nematic (N\textsubscript{TB}) liquid crystals (LCs), the director (mean molecular orientation) exhibits heliconical structure with nanoscale periodicity. On the mesoscopic scale, N\textsubscript{TB} resembles layered systems (like smectics) without a true mass density wave, where the helical pitch is equivalent to a ``pseudo-layer". 
 We study rheological properties of a N\textsubscript{TB} phase and compare the results with those of an usual SmA phase. 
  Analysing the shear response and adapting a simplified physical model for rheology of defect mediated lamellar systems we measure the pseudo-layer compression elastic constant $B_{eff}$ of N\textsubscript{TB} phase from the measurements of dynamic modulus $G^{*}(\omega)$. It is found that $B_{eff}$ of the N\textsubscript{TB} phase is in the range of $10^{3}-10^{6}$ Pa and it follows a temperature dependence, $B_{eff}\sim (T_{TB}-T)^{2}$ as predicted by the recent coarse-grained elastic theory. Our results show that the structural rheology of N\textsubscript{TB} is strikingly similar to that of the usual smectic LCs although the temperature dependence of $B_{eff}$ is much faster than that of smectic LCs as predicted by the coarse-grained models.
 
\end{abstract}
\preprint{HEP/123-qed}
\maketitle

\begin{figure}
\center\includegraphics[scale=0.45]{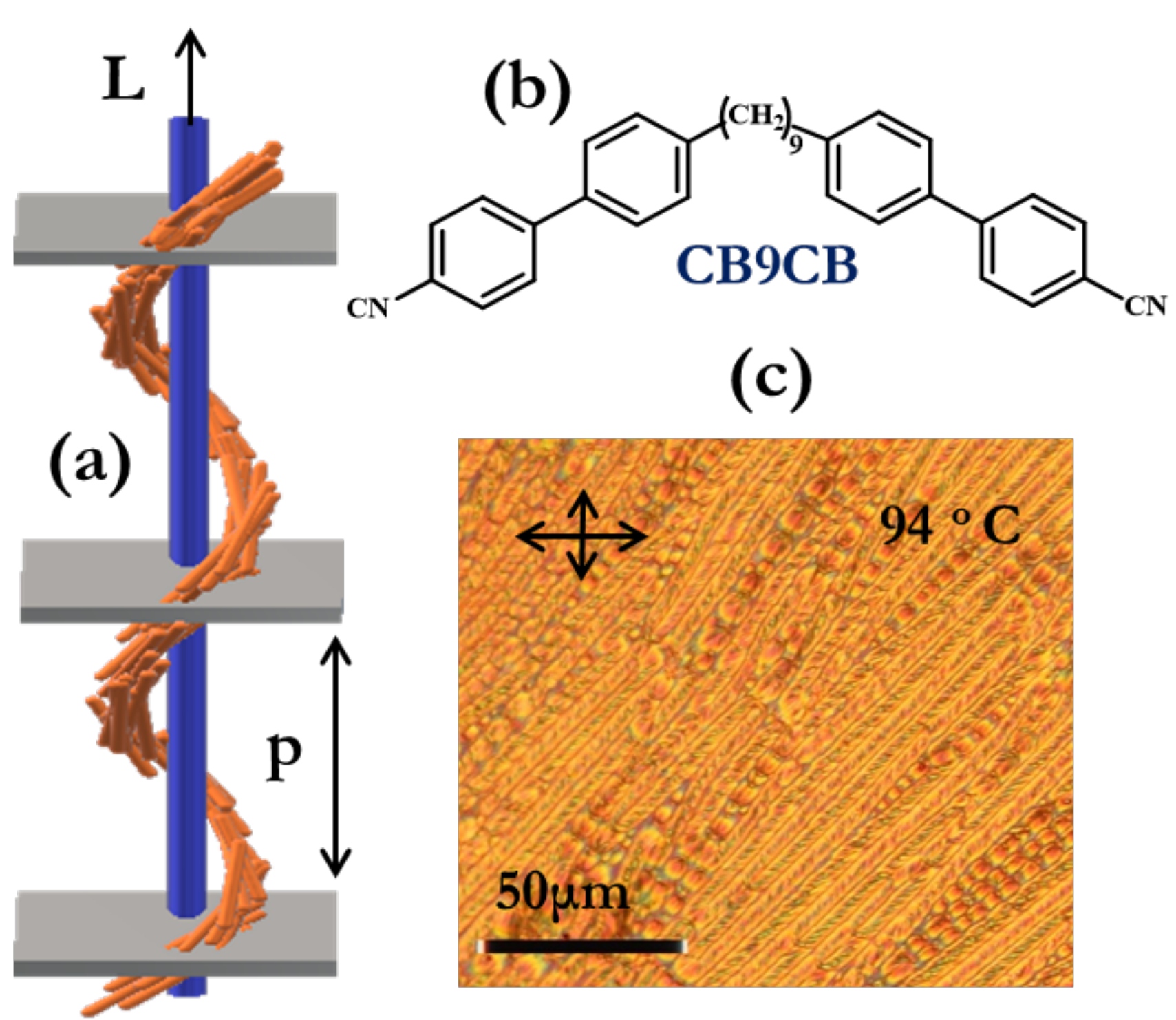}
\caption{(a) Schematic view of  heliconical  molecular orientation of the nematic twsit-bend (N\textsubscript{TB}) phase. $p$ represent helical pitch equivalent to pseudo-layer thickness.  (b) Chemical structure of CB9CB molecules used in the study. (c) Polarising optical microscope texture at 94$^\circ$C in the N\textsubscript{TB} phase. ${\bf L}$ is the macroscopic average orientation of ${\bf\hat{n}}$ over several periods. 
\label{fig:figure1} }
\end{figure}

  
  Experimental discovery of twist-bend nematic (N\textsubscript{TB}) phase in bent-core liquid crystals has created immense interests in liquid crystal community ~\cite {vp,mco,chd,vb,lbe,jzh,gpa} although it was  theoretically predicted much before from different perspectives~\cite{rb1,vl1,vl2,id}. 
  In the N\textsubscript{TB} phase, the director  ${\bf\hat{n}}$ (the mean molecular orientation) exhibits periodic twist and bend deformations forming a conical helix and is tilted with respect to the axis ${\bf L}$  as shown in Fig.\ref{fig:figure1}(a). The typical pitch $p$ of the heliconical structure is of the order of 10 nm, thus comparable to a few molecular length. Commonly, N\textsubscript{TB} phase is observed in odd-membered liquid crystal dimers wherein two mesogenic units are connected through a flexible spacers~\cite{mc,da,rjm}. 
   A fascinating feature of the N\textsubscript{TB} phase is the observation of spontaneous chirality i.e., formation of both left and right handed helical domains even though the constituent molecules are achiral. This leads to several unusual physical properties of N\textsubscript{TB}  phase compared to the conventional nematic phase (N)~\cite{mcl,ntr,nse,sha,par1,cz,rbh}. 
  
  A few coarse-grained theories have been proposed to explain the emergence of N\textsubscript{TB} phase from the high temperature uniform nematic phase. Mayer and Dosov showed that the elastic properties of the N\textsubscript{TB}  phase could be viewed in two different length scales in reference to the pitch length ($p$)~\cite{cm}. When the considered length ($l$) is less than $p$ i.e., $l<p$, the elastic description is similar to that of the usual nematics. On the other hand, when $l>>p$, the elastic description is similar to regular lamellar systems  such as cholesteric and smectic LCs~\cite{cm}. In the latter picture, the thickness of one pitch can be considered as a pseudo-layer and the large-scale elasticity of the N\textsubscript{TB} phase could be described in terms of an effective  pseudo-layer compression elastic constant $B_{eff}$ and an curvature elastic constant $K_{11}^{N}$ and the corresponding  free energy density of N\textsubscript{TB} can be expressed as~\cite{cm}: 
 \begin{equation}
 f_{TB}=\frac{1}{2}B_{eff}\epsilon^2+\frac{1}{2} K^{N}_{11}\left({\frac{1}{R_{1}} +\frac{1}{R_2} }\right)^2
 \end{equation}
 where $\epsilon$ is the pseudo-layer compression and $R_{1}$, $R_{2}$ are the curvatures. Another coarse-grained theory has been developed considering the heli-polar order and their coupling with bend distortions~\cite{par,par1}. Both the theories predicted that the temperature dependence of $B_{eff}$ is much faster than that of the usual SmA LCs.

There have been a very few experimental studies on the measurements of $B_{eff}$ of N\textsubscript{TB} LCs~\cite{ewa,par,sm}. For example, Gorecka \textit{et al.} have measured $B_{eff}$ of CB7CB and some chiral  N\textsubscript{TB} LCs. Their reported values are in the range of a usual SmA LC ($10^6-10^7$ Pa,) and vary inversely with the temperature~\cite{ewa}. Parsouzi \textit{et al.} have reported that $B_{eff}$ is in the range of $10^3-10^4$Pa and it scales as $B_{eff}\sim(T_{TB}-N)^{3/2}$ ~\cite{par}. 
Thus, several orders of magnitude difference in the reported values, measured on two different samples using two different methods and its universal temperature dependence is still an unresolved problem. In this paper we report experimental studies on the rheological properties of a N\textsubscript{TB} LC. We use a novel method for measuring $B_{eff}$ from the dynamic shear modulus $G^{*}(\omega)$. 
We discuss the temperature dependence of $B_{eff}$ and compare with that proposed by the coarse-grained elastic theories of the N\textsubscript{TB} phase.

The LC material  1,$\omega$-bis(4-cyanobiphenyl-4$^{'}$-yl) alkane (CB9CB) studied was synthesised in our laboratory following the procedure reported in Ref.~\cite{da}. It is a cyanobiphenyl-based  dimer with odd number methylene units ($n=9$) in the flexible spacer (Fig.\ref{fig:figure1}(b)).  It exhibits following phase transitions: I $124^\circ$C N $108^\circ$C N\textsubscript{TB} $84^\circ$C Cr. and enantiotropic transition with widest temperature range of N\textsubscript{TB} phase ($22^\circ$C)  among the LCs in the homologous series of CBnCB~\cite{da}. For the purpose of comparison we also studied 8CB (Octylcyano biphenyl) which shows following phase transitions: I $41^\circ$C N $34^\circ$C SmA $22^\circ$C Cr. We used a strain controlled Rheometer (MCR 501, Anton Paar) with a cone-plate measuring system having a plate diameter of 25mm and cone angle of $1^\circ$ for rheological measurements. A Peltier temperature controller was attached with the bottom plate for controlling the temperature with an accuracy of $0.1^\circ$C. A hood was used to cover the measuring plates for uniformity of sample temperature. Temperature dependent viscosity was measured in cooling the sample from the isotropic phase. For measuring the dynamic shear modulus the sample was quenched from the isotropic to the N\textsubscript{TB} phase at the rate of 15$^\circ$C/min.  A total 5g LC was synthesised and about 200mg was used for each rheological measurements. Initially the phase transitions and textures were observed using a polarising optical microscope (Olympus BX51) and a temperature controller (Mettler FP 90). A typical texture of an unaligned sample is shown in Fig.\ref{fig:figure1}(c). It is noticed that the textures of N\textsubscript{TB} are very similar to that of the focal conic textures of usual SmA LCs.

\begin{figure}[!ht]
\center\includegraphics[scale=0.5]{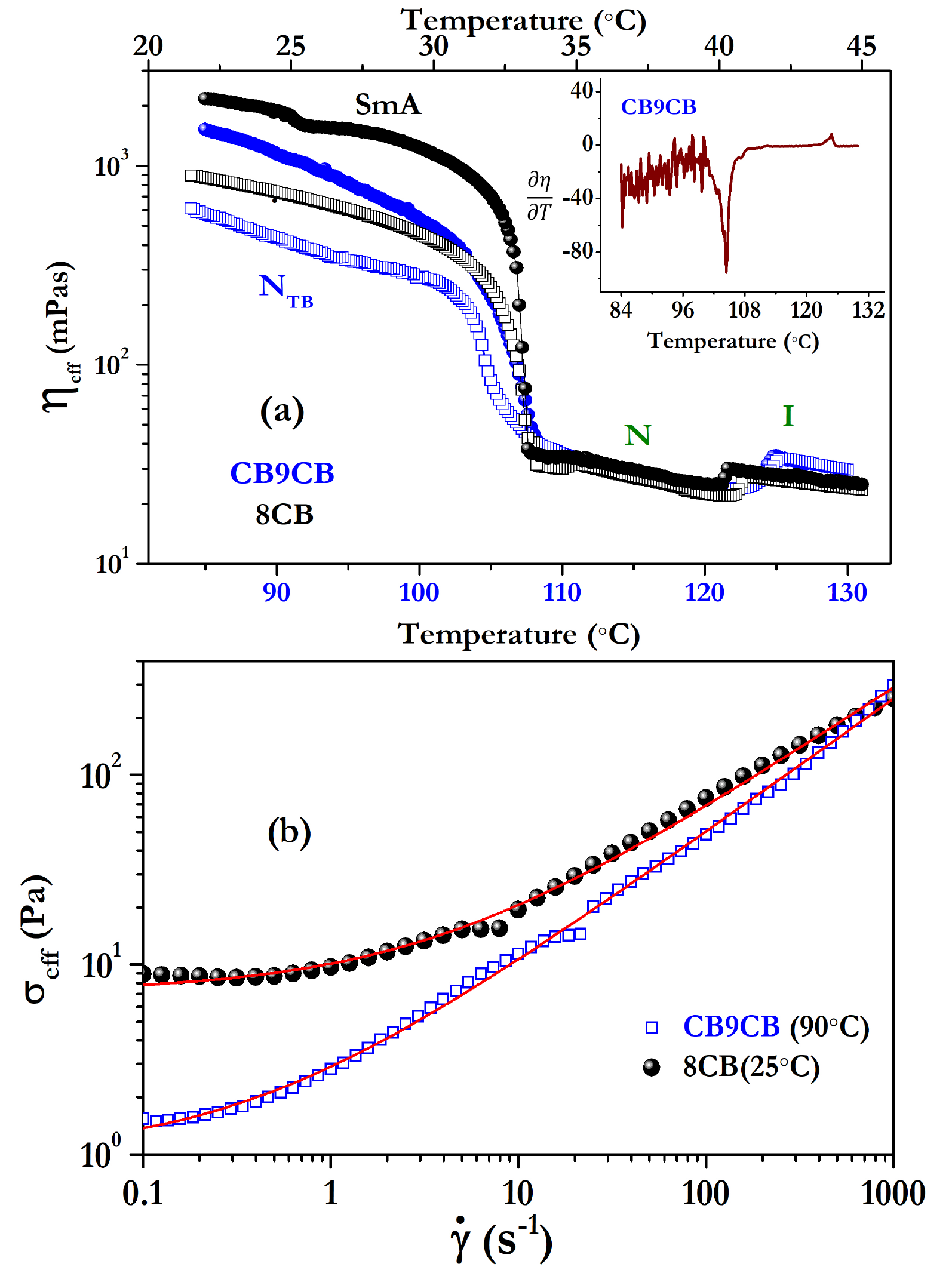}
\caption{(a) Temperature dependent effective viscosity $\eta_{eff}$ of CB9CB (blue squares) and 8CB (black spheres) LCs at two different shear rates namely, $\dot{\gamma}=100$ s\textsuperscript{-1} (squares) and $\dot{\gamma}=10$ s\textsuperscript{-1} (circles). (Inset) Variation of $\frac{\partial\eta_{eff}}{\partial T}$ with temperature $T$ of CB9CB.  (b)  Shear rate  dependent effective shear stress $\sigma_{eff}$ at fixed temperatures. The solid lines are theoretical fits to Eq.(2).
\label{fig:figure2}}
\end{figure}

\begin{figure}[!ht]
\center\includegraphics[scale=0.5]{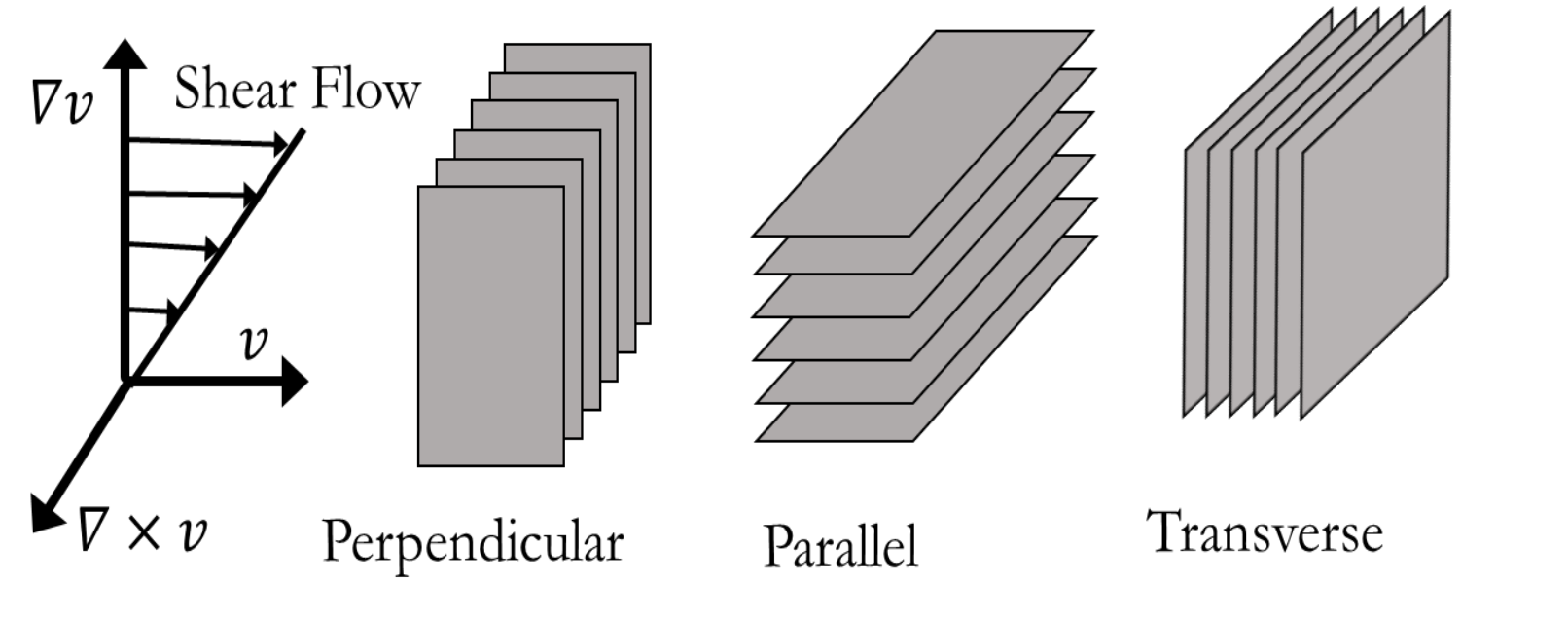}
\caption{Schematic representation of pseudolayer orientations in a shear flow.
\label{fig:figure3}}
\end{figure}

To begin with we measure shear viscosity of CB9CB as a function of temperature at two shear rates ($\dot{\gamma}=$100 and 10 s\textsuperscript{-1}) to identify the phase transition temperatures. Since the orientation of the director with respect to the shear direction usually changes with temperature we define it as effective viscosity $\eta_{eff}$.  As shown in Fig.\ref{fig:figure2}(a) the N to N\textsubscript{TB} transition is identified from the rapid increase (more than two orders of magnitudes) of $\eta_{eff}$ with respect to the N phase. The onset of the N-N\textsubscript{TB} transition ($108^\circ$C) is better seen in the inset of Fig.\ref{fig:figure2}(a). 
We also measured temperature dependent $\eta_{eff}$ of 8CB LC as shown in Fig.\ref{fig:figure2}(a). It is evident that the magnitude and the overall temperature dependence of $\eta_{eff}$ of the two samples are very similar. 
It also presumably indicates that the pretransitional fluctuations and the resulting director dynamics across the N-N\textsubscript{TB} transition are similar to that of N-SmA transition~\cite{crs,jana,praveen}. In analogy with  8CB LC,  three simplest orientations of the pseudo-layers can be considered, wherein the layer normals are parallel to the vorticity ($\nabla\times v$), velocity gradient ($\nabla v$) and flow directions ($v$) as shown schematically in Fig.\ref{fig:figure3}. These are commonly known as perpendicular, parallel and transverse orientations.
The large $\eta_{eff}$ of the N\textsubscript{TB} phase is expected to arise from the transverse orientation of the pseudo-layers similar to those reported in the SmA phase of 8CB LC~\cite{book}. 

The flow curve of N\textsubscript{TB} phase has been studied and compared with that of the SmA phase of 8CB as shown in Fig.\ref{fig:figure2}(b). N\textsubscript{TB} phase shows yield stress similar to that of the SmA phase of 8CB LC. However, the rheology of SmA phase of 8CB is complex as it shows a shear induced structural transition~\cite{pascal}. A small but discontinuous change of  $\sigma_{eff}$ near $\dot{\gamma}=8$s\textsuperscript{-1} in Fig.\ref{fig:figure2}(b) and increase in $\eta_{eff}$ at 24$^{\circ}$C (Fig.\ref{fig:figure2}(a)) could be a signature of such effect. Interestingly similar discontinuity is observed in case of CB9CB at $\dot{\gamma}=22$s\textsuperscript{-1}. Further studies are required to confirm the occurrence  of such transition in CB9CB. Nevertheless, the shear rate dependent shear stress $\sigma_{eff}$ is fitted with the Herschel-Bulkley (HB) model:

\begin{equation}
\sigma_{eff}=\sigma_y+A\dot{\gamma}^n
\end{equation}
where $\sigma_y$ is the yield stress, and $A$ and $n$ are constants. The fit parameters obtained are: $n=0.71$, $\sigma_y=1.0$ Pa, $A=1.9$ for  N\textsubscript{TB} and $n=0.66$, $\sigma_y=7.1$ Pa, $A=2.9$ for SmA phase. The two sets of fit parameters characterising the flow curves of two phases are reasonably close, suggesting they have similar flow behaviour. These two samples have structural similarity namely, the pseudo layer thickness ($\sim$10nm) of twist-bend nematic is closer to the layer thickness of 8CB ($\sim$2nm). Moreover both the samples exhibit focal conic textures. Hence their generic mechanical responses under shear are similar.

\begin{figure}[!ht]
\center\includegraphics[scale=0.58]{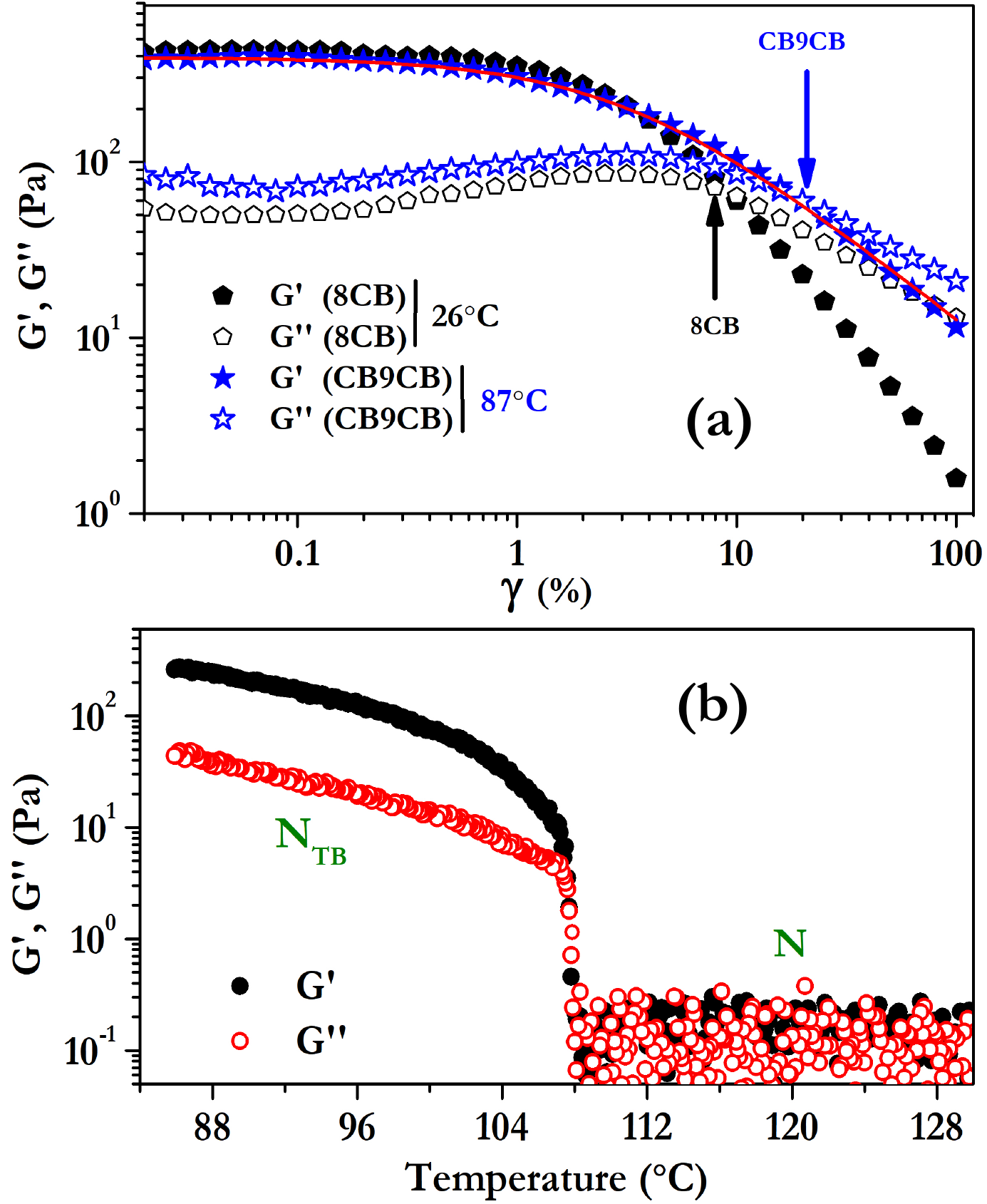}
\caption{  (a) Strain amplitude dependence of the storage $G^{'}$ (solid symbol) and loss  moduli $G^{''}$ (open symbol) for N\textsubscript{TB} (T=$87^\circ$C) and 8CB (T=$26^\circ$C). Solid line is a theoretical fit to Eq.(3). Arrows indicate crossover strains. (b) Temperature dependence of $G^{'}$ (solid symbol) and loss  moduli $G^{''}$ (open symbol) at a fixed strain amplitude $\gamma=0.1\%$. Measurements are performed at frequency $\omega=1$ rad/sec. 
\label{fig:figure4}}
\end{figure}

As a next step, we measure the dynamic modulus $G^{*}(\omega)=G^{'}(\omega)+iG^{''}(\omega)$. The regime of linear viscoelasticity of N\textsubscript{TB} is determined by performing oscillatory measurements in which the strain amplitude is varies from $\gamma=0.01\%$ to  $\gamma=100\%$ at a fixed frequency $\omega=1$ rad/sec.  The strain amplitude dependence can be described by the empirical relation~\cite{rh}, 
\begin{equation}
G^{'}(\omega,\gamma)=\frac{G^{'}(\omega,0)}{1+\gamma/\gamma_c}
\end{equation}
where $\gamma_c$ is the critical strain amplitude.
 As shown in Fig.\ref{fig:figure4}(a) the critical strain amplitude $\gamma_c=3.3\%$ and modulus $G^{'}(1,0)=391$ Pa, setting the upper limit of linear viscoelastic regime. 
For comparing we also measured $G^{*}$ of the SmA phase of 8CB and shown in Fig.\ref{fig:figure4}(a). It is interesting to note that not only the strain dependence but also the magnitudes of the shear moduli of the N\textsubscript{TB} and SmA are comparable, indicating they have common structural origin. 
Further we have measured the temperature dependence of dynamic moduli of N\textsubscript{TB} at a fixed shear amplitude $\gamma=0.1\%$ and observed that $G^{'}>G^{''}$ in the entire N\textsubscript{TB} phase. This is remarkably similar to that is observed in smectics with true mass density wave. Hence, the shear response of the N\textsubscript{TB} phase can be discussed in analogy with the rheological responses of usual SmA liquid crystals~\cite{rh}. Like SmA, N\textsubscript{TB} is solid like in one direction and liquid like in other two directions. Three simplest orientations of the pseudo-layers are considered as shown in Fig.\ref{fig:figure3}. In perpendicular and parallel orientations the pseudo-layers  can slide past each other easily and the N\textsubscript{TB} phase behaves like a liquid.  In transverse orientation the shear tends to change the pseudo-later spacing. As a result of which 
 N\textsubscript{TB} phase shows a viscoelastic solid-like behaviour consequently, $G^{'}>G^{''}$. 
 

\begin{figure}[htbp]
\includegraphics[scale=0.58]{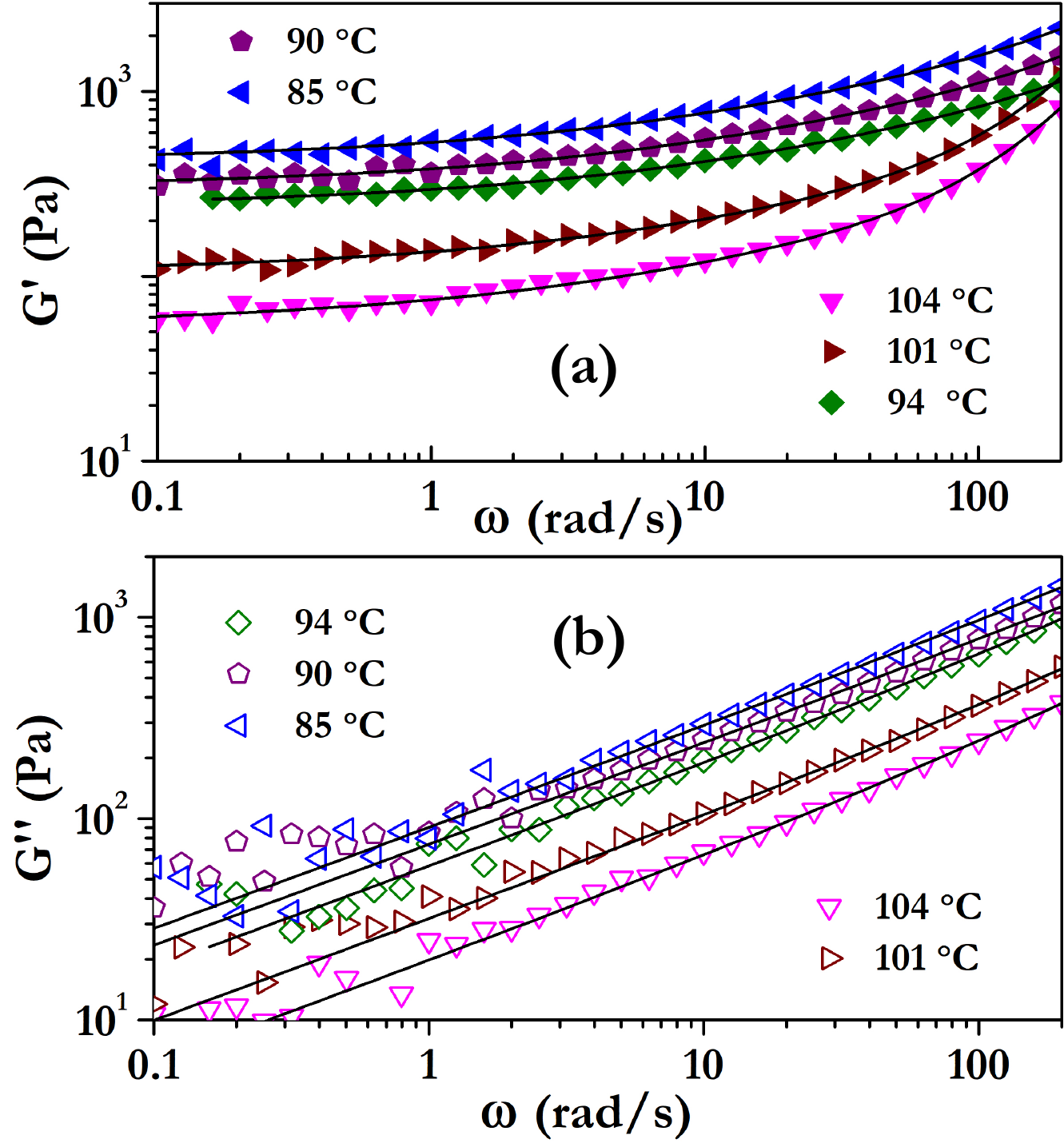}
\center\caption{ Frequency dependence of storage $G^{'}$ (solid symbols) and loss $G^{''}$ (open symbols) moduli at a few representative temperatures. Solid lines are theoretical fits of Eq.(4) and Eq.(5) to $G^{'}$ and $G^{''}$, respectively. 
 \label{fig:figure5}}
\end{figure}


\begin{figure}[htbp]
\center\includegraphics[scale=0.55]{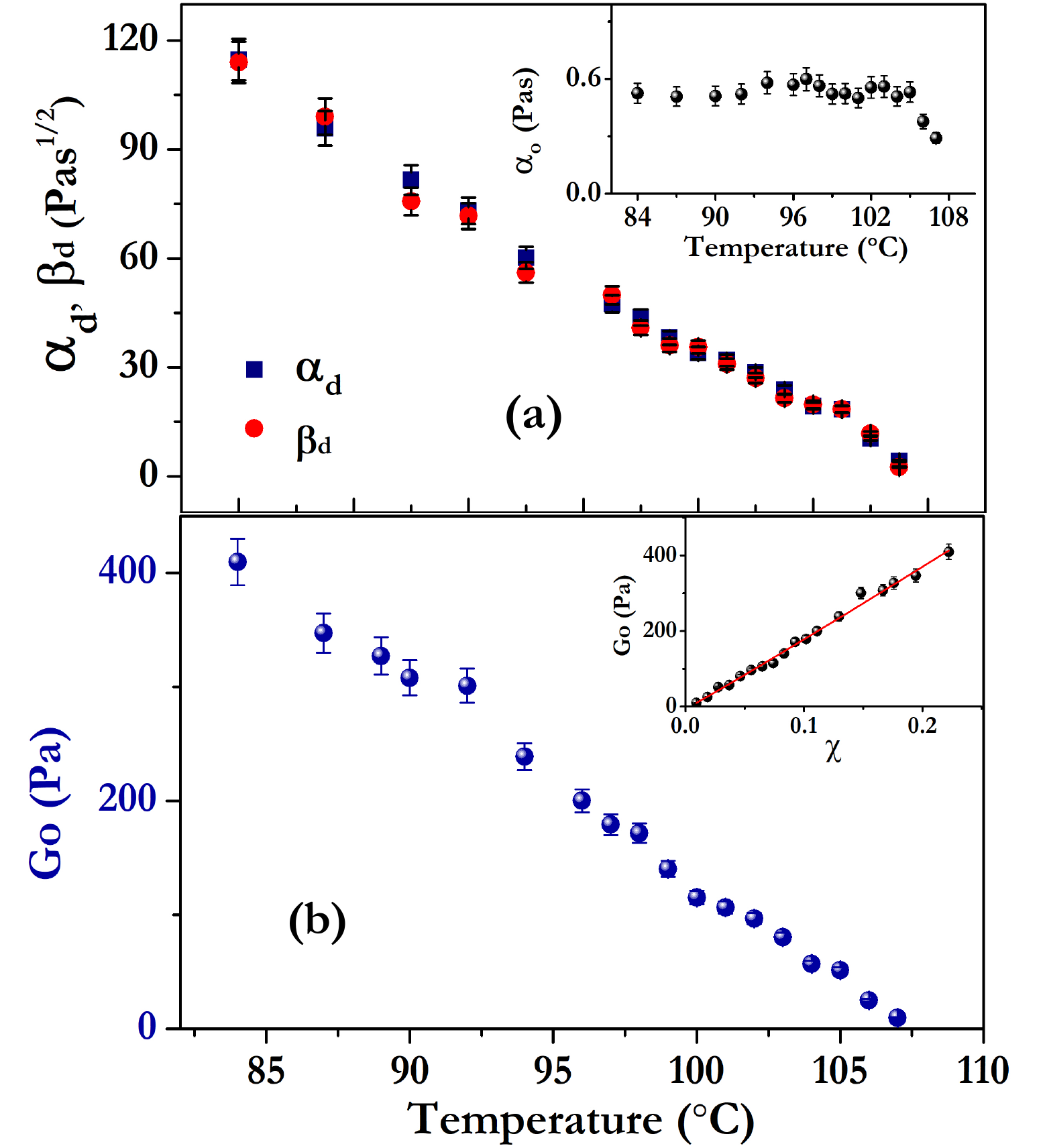}
\caption{(a) Temperature dependence of fit parameters $\beta_d$ (red spheres) and $\alpha_d$ (navy blue squares). (Inset) Temperature dependence of $\alpha_0$. (b) Temperature dependence of fit parameter $G_0$. (Inset) Solid line shows the fit result $G_0\sim\chi$, where the reduced temperature $\chi=(T_{TB}-T)/T_{TB}$.
\label{fig:figure6}}
\end{figure}

\begin{figure}[htbp]
\center\includegraphics[scale=0.52]{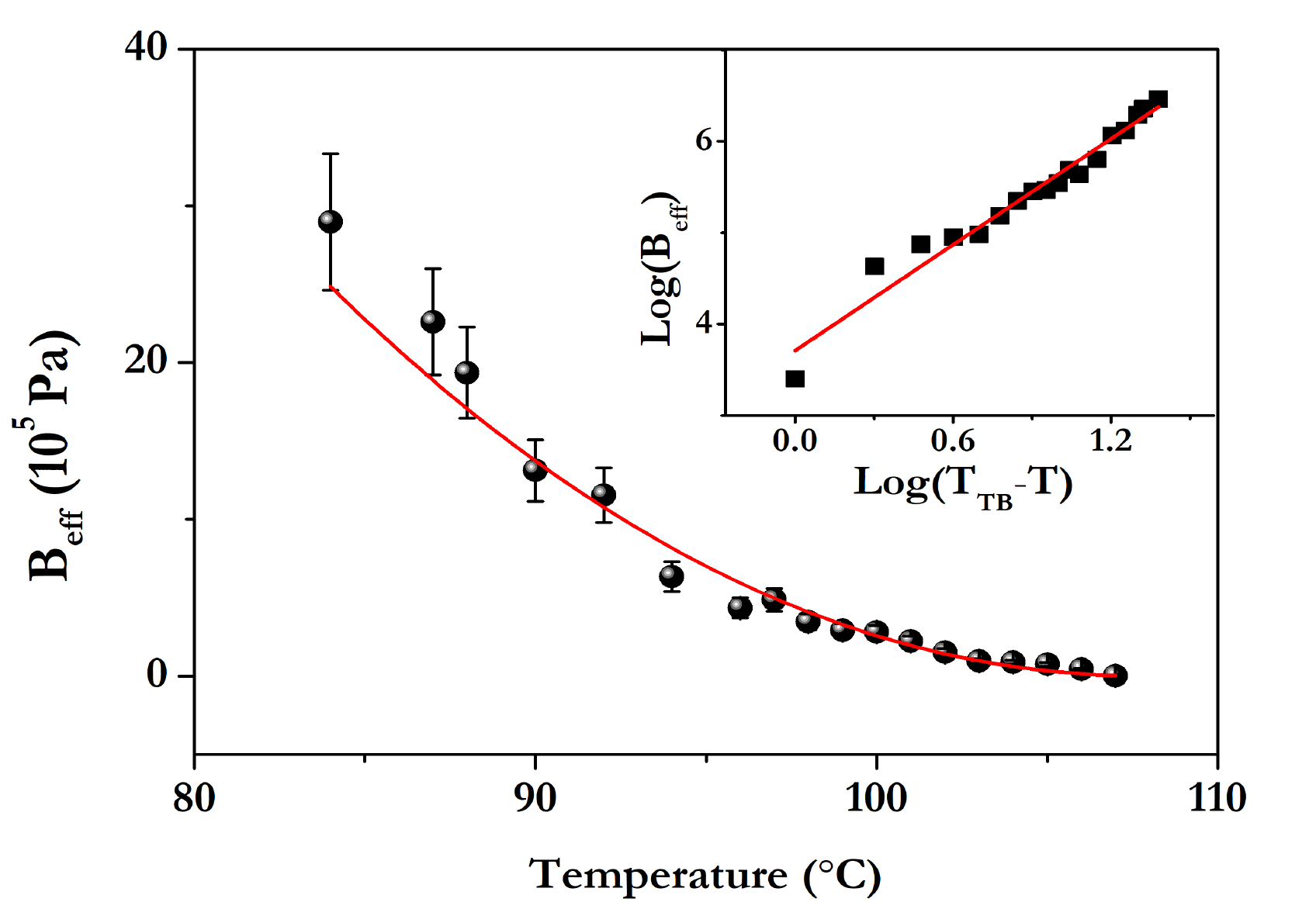}
\caption{Temperature dependence of effective elastic compressional modulus $B_{eff}$. The solid line is theoretical fit to the equation $B_{eff}\sim (T_{TB}-T)^{\alpha}$, where $\alpha=2.0\pm0.1$. (Inset) Log-Log scale.
\label{fig:figure7}}
\end{figure}

The observed solid-like response of the N\textsubscript{TB} phase can be explained based on a simple physical model described for defect mediated cholesteric and smectic LCs. According to this model the storage modulus can be expressed as~\cite{lr}
\begin{equation}
G^{'}(\omega)=G_0+\beta_d\omega^{1/2}+\beta_0\omega^2
\end{equation}
The first term $G_0$ arising from the elasticity of the static defects~\cite{rb}. The second term $\beta_d\omega^{1/2}$ arises from the regions of misaligned pseudo-layers in the sample~\cite{anu}. The last term results from the regions of the sample where the pseudo-layers are parallel to the shear direction. The proportionality constants are  given by $\beta_0\simeq \eta(\gamma_1/K)(p/4\pi)^2$ and $\beta_d=(\pi/24\sqrt{2})\sqrt{(B_{eff}\eta)}$ where $\gamma_1$ is the rotational viscosity, $\eta$ is the effective viscosity and $B_{eff}$ is the compression elastic modulus~\cite{anu}. In particular, $\beta_d$ describes the response of the lamellar regions with the layer normal oriented such that strain involves layer compression.
Following the similar arguments the loss modulus can be written as~\cite{lr}
\begin{equation}
G^{''}(\omega)=\alpha_d\omega^{1/2}+\alpha_0\omega
\end{equation}
where the first and the second terms arise from the  misaligned parts of the samples and the Maxwell-fluid type contribution, respectively.
Thus, by measuring $\beta_d$ and $\alpha_0$ from the disoriented sample, we can estimate $B_{eff}$. In order to measure these parameters at different temperatures we quenched the sample directly from the isotropic to the N\textsubscript{TB} phase so as to obtain mostly disoriented sample. Figure \ref{fig:figure5} shows some representative plots of $G^{'}(\omega)$ and $G^{''}({\omega})$ at different temperatures in the N\textsubscript{TB} phase. The  parameters obtained by fitting Eq.(4) and Eq.(5) at different temperatures are shown in Fig.\ref{fig:figure6}. The fit parameter $\beta_0$ is found to be very small ($10^{-3}$) and does not vary considerably with temperature. It is noted that $\beta_d\simeq\alpha_d$ as expected theoretically and both increases with decreasing temperature. Assuming the whole sample is in the disorientated state, the temperature dependence of $B_{eff}$ can be expressed as
\begin{equation}
B_{eff}=\frac{\beta_d^2}{\alpha_0} \left( \frac{24\sqrt{2}}{\pi} \right)^2
\end{equation}
 Figure \ref{fig:figure7} shows the temperature dependence of calculated $B_{eff}$ in the N\textsubscript{TB} phase. Just below the N-N\textsubscript{TB} transition $B_{eff}$ is relatively smaller; $1.5\times10^3$Pa (T=105$^\circ$C) and it increases rapidly  to $2\times10^6$ Pa (T=85$^\circ$C). The latter number is only one order of magnitude lower that the typical layer compression modulus of SmA phase of 8CB~\cite{mb}. So far experimentally $B_{eff}$ of a very few N\textsubscript{TB} LCs have been measured. Gorecka \textit{et al.} measured temperature dependence of $B_{eff}$ of a few chiral twist bend nematic LCs including CB7CB using atomic force microscopy technique and reported that $B_{eff}$ is in the range of $10^{6}-10^{7}$ Pa, comparable to an ordinary SmA LCs.  Using a dynamic light scattering technique, Parsouzi \textit{et al.} reported that $B_{eff}$ of N\textsubscript{TB} phase of a LC made of multicomponent mixture is in the range of $10^{3}-10^{4}$ Pa,  which is almost three orders of magnitude smaller than the ordinary SmA LCs~\cite{sm,par}. Our experiment shows a wide variation of $B_{eff}$, covering both the ranges. Such wide variation of $B_{eff}$ could partly be attributed to the increase in the cone angle $\theta$ with decreasing temperature. Based on a crude model $B_{eff}=K_2 (2\pi/p)^2 \sin^4\theta$~\cite{sm}, where $\theta$ increases with decreasing temperature~\cite{cz,cg}. Considering physical parameters for CB7CB such as:  $p=10$nm~\cite{cz} and $\theta=10^{\circ}$~\cite{cg} (2$^\circ$C below N-N\textsubscript{TB} transition); $p=8$nm~\cite{cz} and $\theta=33^{\circ}$~\cite{cg} (25$^\circ$C below the transition) and $K_2=3$pN, the calculated $B_{eff}$ near two limiting temperatures are given by $1.1\times10^3$ and $1.6\times10^5$ Pa.  The calculated $B_{eff}$ close to the transition agrees reasonably well with our experiments but it is smaller by one order of magnitude at far below the transition. This assessment suggests that mere increase in the cone angle with decreasing temperature can enhance $B_{eff}$ by almost two orders of magnitude.

The temperature dependence of $B_{eff}$ have been predicted theoretically using coarse-grained theoretical models. 
 In analogy with  SmA\textsuperscript{*}, Meyer and Dosov defined smectic-like effective pseudo-layer compression and bending  elastic constants ($K_{33}^N$)~\cite{cm}. Assuming $K_{33}^N<0$, they predicted $B_{eff}\sim (T_{TB}-T)^2$. Parsouzi \textit{et al.} have developed another model accounting the helical polarisation field and their coupling with the bend distortion of the director~\cite{par}. Considering a small variation in the pseudo-layer spacing and resulting changes in the cone angle and polar order, the theory predicts that there are two regimes of $B_{eff}$. For temperature (T) sufficiently close to the T\textsubscript{TB}, $B_{eff}\sim(T_{TB}-T)^{3}$, whereas for T sufficiently below T\textsubscript{TB}, the theory gives $B_{eff}\sim(T_{TB}-T)^{3/2}$. Experimentally they found (in a mixture exhibiting N\textsubscript{TB} phase) that within a relatively small temperature range ($\sim 6^{\circ}$), $B_{eff}$ scales as $(T_{TB}-T)^{3/2}$. 
 To get an estimation of the scaling exponent of $B_{eff}$, we fit our data to the equation $B_{eff}\sim(T_{TB}-T)^{\alpha}$, with $\alpha$ as a fit parameter as shown in Fig.\ref{fig:figure7}. Inset shows the variation in Log-Log scale. We obtain $\alpha=2.0\pm0.1$, which is equal  to the scaling exponent predicted by the ``negative elasticity'' model of Meyer and Dosov~\cite{cm}. 


The temperature dependence of $G_0$ as shown in Fig.\ref{fig:figure6}(b) indicates that the static defects contribute to the mechanical response of the N\textsubscript{TB} phase similar to that of the SmA phase. Inset of Fig.\ref{fig:figure6}(b) shows that it scales with temperature as $G_0\sim\chi$, where $\chi=(T_{TB}-T)/T_{TB}$ is the reduced temperature. This is slightly faster than that in case of SmA ($\chi^{0.7}$)\cite{sfu} as expected in view of the fact that $B$ scales much faster with temperature in the N\textsubscript{TB} phase.

 Three remarks are in order. Firstly, in estimating $\beta_d$ it has been assumed that the pseudo-layers are disoriented in the whole sample. This assumption is reasonable as the sample was quenched from the isotropic phase. Nevertheless if the sample is partly oriented we do not expect an order of magnitude reduction of $\beta_d$ and consequently $B_{eff}$. Following the same procedure $B_{eff}$ of 8CB is measured at a few temperatures which agrees well with that was reported~\cite{mb}. 
 Secondly, the experimental results indicate that there are defects in our system whose contribution to the elasticity increases linearly with decreasing temperature. This might indicate a type-II N\textsubscript{TB} having twist grain boundary (TGB) like structure as predicted by the coarse-grained model~\cite{cm}. However, at this point we can say that more experimental investigations are required to draw any unambiguous conclusion. Lastly, it may be mentioned that though structurally N\textsubscript{TB} is closer to cholesterics but the mechanical responses of N\textsubscript{TB} and smectics are very much alike. This could be attributed to the fact that the thickness of pseudo layers in N\textsubscript{TB} is much closer to smectic layer than the pitch of the usual cholesterics. 


In this work we have presented rheological properties of a N\textsubscript{TB}  LC.   The structural rheology of N\textsubscript{TB} phase is found to be remarkably similar to that of the usual SmA phase of calamitic liquid crystals. Our measurements reveal that in spite of the absence of true mass density wave,  N\textsubscript{TB} LCs are viscoelastic solids similar to many defect mediated lamellar systems.  We found that $B_{eff}$ is relatively softer near the N-N\textsubscript{TB} transition but increases with decreasing temperature to three orders of magnitude more at a much faster rate than the usual SmA LCs. The temperature dependence of $B_{eff}$ agrees well with the prediction of the coarse-grained elastic theories.  
 Thus our results provide a valuable test of the validity of the proposed theoretical models. The experiments also offer new perspectives of N\textsubscript{TB} LCs and open unexplored aspects of rheology of nematic LCs with nanoscale modulation of the director. \\


\textbf{Acknowledgments}: SD acknowledges financial support from SERB (Ref. No:CRG/2019/000425) and DST-FIST-II, School of Physics. PK acknowledges financial support from MUT UGB 22-760. PK acknowledges UGC-CSIR for fellowship.


\begin{thebibliography}{99}
\bibitem{vp}V. P. Panov, M. Nagaraj, J. K. Vij, Yu. P. Panarin, A. Kohlmeier, M. G. Tamba, R. A. Lewis, and G. H. Mehl \textcolor{blue}{Phys. Rev. Lett., \textbf {105}, 167801 (2010).} 

\bibitem{mco} M. Copic,  \textcolor {blue} { Proc. Natl. Acad. Sci. USA,  \textbf {110},  15931 (2013).}

\bibitem{chd} D. Chen, J. H. Porada, J. B. Hooper, A. Klittnick, Y. Shena, M. R. Tuchbanda, E. Korblova, D. Bedrov, D. M. Walba, M. A. Glaser, J. E. Maclennana, and N. A. Clark, \textcolor {blue} { Proc. Natl. Acad. Sci. USA,  \textbf {110}(40),  15855 (2013).}

\bibitem{vb}V. Borshch, Y.-K. Kim, J. Xiang, M Gao, A Jákli, V. P. Panov, J. K. Vij, C. T. Imrie, M. G. Tamba, G. H. Mehl, and O. D. Lavrentovich,  \textcolor {blue} {Nat. Commun., \textbf {4}, 2635 (2013).}

\bibitem{lbe} L. Beguin, J. W. Emsley, Moreno Lelli, A. Lesage, G. R. Luckhurst, B. A. Timimi, and H. Zimmermann, \textcolor {blue} {J. Chem. Phys. B \textbf {116}, 7940 (2012).}

\bibitem{gpa} G. Pajak, L. Longa, and A. Chrzanowska,  \textcolor  {blue} {Proc. Natl. Acad. Sci. USA,  \textbf {115}(40),  E10303 (2018).}

\bibitem{jzh}J. Zhou, W. Tang, Y. Arakawa, H. Tsuji and S. Aya, \textcolor {blue} {Phys. Chem. Chem. Phys., DOI: 10.1039/C9CP06861A (2020).}

\bibitem{rb1} R. B. Meyer,, in Molecular Fluids, ed. R. Balian and G. Weill, Les Houches Summer School in Theoretical Physics, Gordon and Breach, New York, vol. XXV-1973, pp.273-373 (1976). 

\bibitem{vl1} V. L. Lorman and B. Mettout,  \textcolor {blue} {Phys. Rev. Lett., \textbf{82}, 940 (1999)}.

\bibitem{vl2} V. L. Lorman and B. Mettout,   \textcolor {blue} {Phys. Rev. E \textbf{69}, 061710 (2004)}.

\bibitem{id} I. Dosov, \textcolor {blue} {EPL \textbf{56}, 247 (2001)}.


\bibitem{rjm} R. J.  Mandle, E. J. Davis, C. T. Archbold, S. J. Cowling and J. W. Goodby, \textcolor {blue} {J. Mater Chem. C, \textbf {2}, 556 (2014).}


\bibitem{mc} M. Cestari, S. Diez-Berart, D. A. Dunmur, A. Ferrarini, M. R. de la Fuente, D. J. B. Jackson, D. O. Lopez, G. R. Luckhurst, M. A. Perez-Jubindo, R. M. Richardson, J. Salud, B. A. Timimi, and H. Zimmermann \textcolor {blue} {Phys. Rev. E \textbf{84}, 031704 (2011)}.

\bibitem{da} D.A. Paterson, J. P. Abberley, W. TA. Harrison, J. MD Storey and C. T. Imrie, \textcolor {blue} {Liq. Cryst. \textbf{44}, 127 (2017)}.

\bibitem{mcl} C. Meyer, G. R. Luckhurst, and I.  Dozov,  \textcolor {blue} {Phys. Rev. Lett., \textbf {111}, 067801 (2013).}

\bibitem{nse} N. Sebastian, B. Robles-Hernandez, S. Diez-Berart, J. Salud, G. R. Luckhurst, D. A. Dunmur, D. O. Lpez  and M. R. de la Fuente, \textcolor {blue} {Liq. Cryst. \textbf {44}(1),  177 (2017).}

\bibitem{ntr} N. Trbojevic, D. J. Read, and M. Nagaraj, \textcolor {blue} {Phys. Rev. E \textbf {96},  052703 (2017).}

\bibitem{sha} S. A. Pardaev, S. M. Shamid, M. G. Tamba, C. Welch, G. H. Mehl, J. T. Gleeson, D. W. Allender, J. V. Selinger, B. Ellman, A. Jakli and S. Sprunt, \textcolor {blue} {Soft Matter \textbf {12},  4472 (2016).}

\bibitem{cz} C. Zhu, M. R. Tuchband, A. Young, Min Shuai, A. Scarbrough, D. M. Walba, J. E. Maclennan, C. Wang, A. Hexemer, and N. A. Clark, \textcolor {blue} {Phys. Rev. Lett. \textbf {116}, 147803 (2016)}.

\bibitem{par1} Z. Parsouzi, S. M. Shamid, V. Borshch, K. Challa, A. R. Baldwin, M. G. Tamba, C. Welch, G. H. Mehl, J. T. Gleeson, A. Jakli, O. D. Lavrentovich, D. W. Allender, J. V. Selinger and S. Sprunt, \textcolor {blue} {Phys. Rev. X \textbf {6}, 021041 (2016)}.

\bibitem{rbh} B. Robles-Hernandez, N. Sebastian, M. R. de la Fuente, D. O. Lopez, S. Diez-Berart, J. Salud, M. Blanca Ros, D. A. Dunmur, G. R. Luckhurst, and B. A. Timimi, \textcolor {blue} {Phys. Rev. E \textbf {92}, 062505 (2015)}.


\bibitem{cm} C. Meyer and I. Dosov, \textcolor {blue} {Soft Matter \textbf{12}, 574 (2016)}.

\bibitem{par} Z. Parsouzi, S. A. Pardaev, C. Welch, Z. Ahmed, G. H. Mehl, A. R. Baldwin, J. T. Gleeson, O. D. Lavrentovich, D. W. Allender, J. V. Selinger, A. Jakli and S. Sprunt, \textcolor {blue} {Phys. Chem. Chem. Phys.,  \textbf{18}, 31645 (2016)}.


\bibitem{ewa}E. Gorecka,N. Vaupotic, A. Zep, D. Pociecha, J. Yoshioka, J. Yamamoto and H. Takezoe, \textcolor {blue} {Angew. Chem. Int. Ed. \textbf{54}, 10155 (2015)}.


\bibitem{sm} S. M. Salili, C. Kim, S. Sprunt, J. T. Gleeson, O. Parri and A. Jakli,  \textcolor {blue} {RSC Adv.,  \textbf{4}, 57419 (2014)}.



\bibitem{crs} C. R. Safinya, E. B. Sirota and R. J. Plano, \textcolor {blue} {Phys. Rev. Lett.,  \textbf{66}, 1986 (1991)}.

\bibitem{jana} J. Ananthaiah, M. Rajeswari, V.S.S. Sastry and S. Dhara, \textcolor {blue} {Eur. Phys. J. E  \textbf{34}, 74 (2018)}.

\bibitem{praveen} M. Praveen Kumar, D. Venkata Sai, and Surajit Dhara, \textcolor {blue} {Phys. Rev. E  \textbf{98}, 062701 (2011).}

\bibitem{book} R. G. Larson, The Structure and Rheology of Complex Fluids, Oxford University Press, New York (1999) pp 480.

\bibitem{pascal} P. Panizza, P. Archambault and D. Roux, \textcolor {blue} {J. Phys. II France,  \textbf{5}, 303 (1995)}.

\bibitem{rh} R. H. Colby, C. K. Ober, J. R. Gillmor, R. W. Connelly, T. Doung, G. Galli and M. Laus,  \textcolor {blue} {Rheol. Acta,  \textbf{36}, 498 (1997)}.

\bibitem{lr} L. Ramos, M. Zapotocky, T. C. Lubensky and D. A. Weitz, \textcolor {blue} {Phys. Rev. E  \textbf{66}, 031711-1 (2002)}.

\bibitem{rb} R. Bandyopadhyay, D. Liang, R. H. Colby, J. L. Harden, and Robert L. Leheny, \textcolor {blue} {Phys. Rev. Lett. \textbf {94}, 107801 (2005)}.


\bibitem{anu} K. Kawasaki and A. Anuki, \textcolor {blue} {Phys. Rev. A  \textbf{42}, 3664 (1990)}.

\bibitem{mb} M. Benzekri, J. P. Marcerou, H. T. Nguyen and J. C. Roullion, \textcolor {blue} {Phys. Rev. B \textbf {41}, 9032 (1990)}.


\bibitem{cg} C. Meyer, G. R. Luckhurst and I. Dosov, \textcolor {blue} {J. Mater. Chem. \textbf {3}, 318 (2015)}.

\bibitem{jan} J.A.N. Zasadzinski, \textcolor {blue} {J. Phys. (Les Ulis, Fr.) \textbf {51}, 747 (1990)}.

\bibitem{mkl} M. Kleman, \textcolor {blue} {Rep. Prog. Phys. \textbf {52}, 555 (1989)}.

\bibitem{rll} R. L. Leheny, S. Park, R. J. Birgeneau, J.-L. Gallani, C. W. Garland, and G. S. Iannacchione, \textcolor {blue} {Phys. Rev. E \textbf {67}, 011708 (2003)}.

\bibitem{sfu} S. Fujii, S. Komura, Y. Ishii,  C.-Y.D. Lu,  \textcolor {blue} {J. Phys.: Condens. Matter \textbf{23}, 235105 (2011).}

\end {thebibliography}
\end{document}